\documentclass[aps,pra,twocolumn,showpacs,letterpaper,superscriptaddress]{revtex4-1}

\usepackage[colorlinks=true, citecolor=blue, linkcolor=blue, urlcolor=blue]{hyperref}

\usepackage{graphicx,dcolumn,longtable,epsfig}
\usepackage[usenames]{color}
\usepackage{amssymb}
\usepackage{amsmath}
\usepackage{epstopdf}
\usepackage{bm}
\usepackage{footnote}
\usepackage{float}
\usepackage{subfigure}
\usepackage{color}

\input epsf.tex

\def\bea{\begin{eqnarray}}
\def\eea{\end{eqnarray}}
\def\be{\begin{equation}}
\def\ee{\end{equation}}

\begin{document}

\title{Equilibrium Phases of Tilted Dipolar Lattice Bosons}
\author{ C. Zhang }
\affiliation{Homer L. Dodge Department of Physics and Astronomy,
The University of Oklahoma, Norman, Oklahoma ,73019, USA}
\author{ A.~Safavi-Naini}
\affiliation{JILA , NIST and Department of Physics, University of Colorado, 440 UCB, Boulder, CO 80309, USA}
\author{Ana Maria Rey}
\affiliation{JILA , NIST and Department of Physics, University of Colorado, 440 UCB, Boulder, CO 80309, USA}
\author{B. Capogrosso-Sansone}
\affiliation{Homer L. Dodge Department of Physics and Astronomy, The University of Oklahoma, Norman, Oklahoma ,73019, USA} 
\affiliation{Department of Physics, Clark University, Worcester, Massachusetts 01610}
\begin{abstract}
The recent advances in creating nearly degenerate quantum dipolar gases in optical lattices are opening the doors for the exploration of equilibrium physics of quantum systems with anisotropic and long-range dipolar interactions. In this paper we study the zero- and finite-temperature phase diagrams of a system of hard-core dipolar bosons at half-filling, trapped in a two-dimensional optical lattice. The dipoles are aligned parallel to one another and tilted out of the optical lattice plane by means of an external electric field. At zero-temperature, the system is a superfluid at all tilt angles $\theta$ provided that the strength of dipolar interaction is below a critical value $V_c(\theta)$. Upon increasing the interaction strength while keeping $\theta$ fixed, the superfluid phase is destabilized in favor of a checkerboard or a stripe solid depending on the tilt angle. We explore the nature of the phase transition between the two solid phases and find evidence of a micro-emulsion phase, following the Spivak-Kivelson scenario, separating these two solid phases. Additionally, we study the stability of these quantum phases against thermal fluctuations and find that the stripe solid is the most robust, making it the best candidate for experimental observation.

\end{abstract}

\maketitle

\section{Introduction}

Experimental progress in trapping and controlling ultra cold atoms and molecules has led to the observation of magnetic and electric dipolar interactions in a variety of systems~\cite{Griesmaier:2005fd, Lahaye:2007fo, Ni:2008hma, Ospelkaus:2008fu, Chotia:2012ht, Wu:2012gf, Yan:2013fn, Takekoshi:2014ku, Lu:2012bd, Aikawa:2014kx, Aikawa:2014if, Park:2015dj, Frisch:2015wu}. These systems can be used as quantum simulators to study quantum transport and dynamical and equilibrium properties of models featuring long-range and anisotropic dipolar interactions. 
Long-range dipolar interactions have been predicted to stabilize a plethora of exotic quantum phases such as p-wave superfluids, superfluids of multimers, charge density waves, stripe solids, and supersolids. Moreover, dipolar interactions play an important role in many models of strongly correlated systems, excitons with spatially separated electrons
and holes~\cite{Lerner:1981vg, Filinov:2006ki, Fedorov:2014bu}, and frustrated quantum magnets~\cite{Sachdev:2008fr, Lacroix:2011wn, Wall:2014vj}. 

The recent success in creating a gas of polar molecules in an optical lattice is the first step towards the realization of a low-entropy initial state which is a key ingredient for any quantum simulator~\cite{Yan:2013fn,  Hazzard:2014bx}. The ability to change the alignment of the dipole moment by applying an external field has sparked numerous studies on the many-body phases of dipolar gases. Many-body bosonic and fermionic dipoles with dipoles aligned perpendicular to the lattice or parallel to the lattice have been studied extensively~\cite{Yi:2007dt, Menotti:2007fv, Pollet:2010cc, CapogrossoSansone:2010em, Ohgoe:2012cx, CapogrossoSansone:2011eq, CapogrossoSansone:2011jk, SafaviNaini:2013dh, SafaviNaini:2014kc}. However, the theoretical and numerical effort to study dipolar gases with dipoles aligned at arbitrary tilt angles has either focused on continuous systems~\cite{Yamaguchi:2010fn, Sun:2010go, Parish:2012gq, Macia:2012ci, Macia:2012bq, Ruggeri:2013vc, Macia:2014je} or has employed non-exact methods such as functional renormalization group~\cite{Bhongale:2012fb}, mean field theory and variational approaches~\cite{Goral:2002hu, Danshita:2009cp, Fedorov:2013fc}. 

In the following we present a systematic study of a system of hard-core, dipolar bosons trapped in a two-dimensional lattice. We treat the alignment of the dipoles as a parameter which can be adjusted via the application of an external field. We use the worm algorithm~\cite{Prokofev:1998gz}, a type of Path Integral Quantum Monte Carlo method, and study the zero- and finite-temperature phase diagrams of the system at \emph{half-filling} as a function of the tilt angle and the strength of dipolar interaction. We find that at low interaction strength the system is in the superfluid state for any value of the tilt angle. Upon increasing the interaction strength the superfluid phase is destabilized in favor of either a checkerboard or a stripe solid depending on the tilt angle. We find evidence of the Spivak-Kivelson scenario~\cite{Ng:1995ca, Spivak:2003ka, Spivak:2004jk, Pollet:2010cc} in the transition region between the two solid phases where an emulsion phase of the checkerboard and stripe solids is observed. Additionally, we study the robustness of these quantum phases against thermal fluctuations and show that the solid phases survive at temperatures higher than the critical temperature for the disappearance of the superfluid phase. In particular, due to the anisotropy of the dipolar interaction, the stripe solid turns out to be the most robust phase, making it the best candidate for experimental observation. We give predictions for actual experimental setups and temperatures required to observe solid phases.

This paper is structured as follows: in section~\ref{sec:sec1} we discuss the Hamiltonian describing the system. In section~\ref{sec:sec2} we present the zero and finite-temperature phase diagrams. In section~\ref{sec:sec3} we discuss how harmonic confinement affects our results. In section~\ref{sec:sec4} we explore possible experimental realizations and provide temperature estimates. Finally, section~\ref{sec:sec5} concludes the paper. 

\label{sec:sec1}\section{Hamiltonian}
\begin{figure}[h]
\includegraphics[trim={4cm 2cm 1cm 2cm}, clip, width=0.5\textwidth]{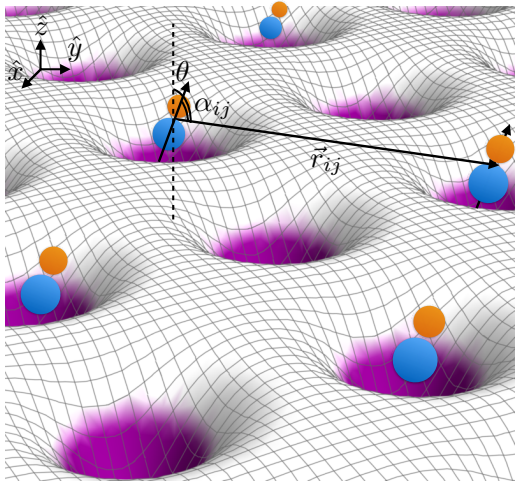}
\caption{(Color online) Schematic representation of the system. Dipoles are trapped in a two-dimensional optical lattice with their dipole moments aligned with the external electric field. The dipole moments make an angle $\theta$ with respect to the axis perpendicular to the plane of the optical lattice. The anisotropic dipolar interaction depends on the angle $\alpha_{ij}$ between the direction of polarization and the relative position of particles.}

\label{fig:fig1}
\end{figure}

We study a system of hardcore, dipolar bosons with induced dipole moment $d$, confined by a two-dimensional optical lattice with lattice constant $a$ and by an external harmonic trap. A schematic representation of this setup, in the absence of the external confinement, is shown in figure~\ref{fig:fig1}. In this system, the dipoles are aligned parallel to each other along the direction of polarization which forms a tilt angle $\theta$ with an axis perpendicular to the plane of the optical lattice. The system is described by the Hamiltonian

\begin{align}
\label{eq:H}
\nonumber H&=-J\sum_{\langle i\,j\rangle }a_i^\dagger a_j+V\sum_{i<j}{\frac{n_{i}n_{j}}{r_{ij}^{3}}(1-3\cos^{2}\alpha_{ij})},\\
& -\sum_i \mu_i  n_i.
\end{align}
where the first term describes the kinetic energy of the system with hopping energy $J$ and the second term is the dipole-dipole interaction with strength $V=d^2/a^3$ and $r_{ij}=\vert \overrightarrow r_i-\overrightarrow r_j\vert$. $\alpha_{ij}$ characterizes the angle between the direction of the polarization and the relative position of the two particles given by $\vec r_{ij}$.  Here, $a_i^\dagger$ ($a_i$) are the bosonic creation (annihilation) operators with the usual commutation relations and $a_i^{\dagger 2}=0$ , $n_i=a_i^\dagger a_i$. We use $\langle \dots \rangle$ to denote nearest neighboring sites. Finally, $\mu_i=\mu-\sum_{\xi=x,y} w_\xi \xi_i^2$, where $w_\xi$ and $\xi_i$ are the strength of harmonic confinement and the coordinate of site $i$ along axis $\xi$, respectively, and $\mu$ is the chemical potential which sets the total number of particles.

\label{sec:sec2}\section{Zero- and Finite-temperature Phase Diagrams}

In this section, we present the zero- and finite-temperature phase diagrams of the system described by Eq.~\ref{eq:H} at half-filling and in the absence of an external harmonic confinement. The chemical potential is set to ensure the half filling condition, $n=N/N_{\rm sites}=0.5$. The effect of the harmonic confinement is discussed in section~\ref{sec:sec3}.  Our results are based on Path Integral Quantum Monte Carlo (QMC) simulations using the worm algorithm (WA). Unless otherwise noted, we simulate systems with spacial dimensions $L\times L$, where $L=24$, 30, 36, and 42, with $N_{\rm sites}=L\times L$. To extract the ground state phase diagram, we work at inverse temperature $\beta=L/J$ which ensures that the system is effectively at zero temperature . 
%$L_\tau=L^z$ is the imaginary time dimension which scales according to the dynamical critical exponent $z$. The inverse temperature is given by $\beta=1/k_BT \propto L_\tau$.
% In the following we set $z=1$ and $\beta=\frac{L}{2J}$.   

\begin{figure}[h]
\includegraphics[trim={1cm 7cm 12cm 1cm}, clip, width=0.5\textwidth]{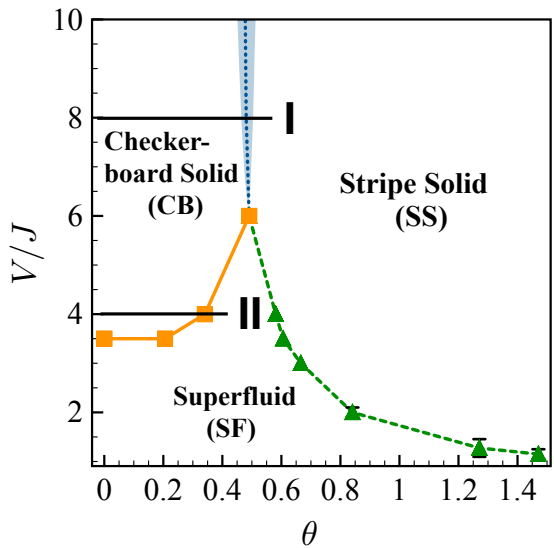}
\caption{(Color online) Zero-temperature phase diagram of the system described by Eq.~\ref{eq:H} as a function of tilt angle $\theta$ and $V/J$, and in the absence of harmonic confinement. The system features a checkerboard solid (CB), a stripe solid (SS), and a superfluid phase (SF). The CB-SS transition follows a Spivak-Kievelson scenario where stripes of one phase are surrounded by the other phase. The shaded region indicates the area where we have observed this micro-emulsion phase.}

\label{fig:fig2}
\end{figure}

The main panel in Figure~\ref{fig:fig2} shows the zero-temperature phase diagram in the $V/J-\theta$ plane which features three phases: a superfluid phase (SF), a checkerboard solid (CB) phase, and a stripe solid (SS) phase. The superfluid phase possesses off-diagonal long-range order and is characterized by finite superfluid stiffness $\rho_{s}$, which can be extracted from simulations by measuring the winding number in space $\rho_{s}=\langle \mathbf{W}^2\rangle /dL^{d-2}\beta$~\cite{Winding}. The diagonal order in the solid phases is characterized by a finite value of structure factor $S(\mathbf{k})=\sum_{\mathbf{r,r'}}  \exp[i \mathbf{k (\mathbf{r-r'})}]\langle n_{\mathbf{r}} n_{\mathbf{r'}} \rangle /N$, where $\mathbf{k}$ is the reciprocal lattice vector. We use $\mathbf{k}=(\pi, \pi)$ and $\mathbf{k}=(0, \pi)$ to identify the CB and SS phases, respectively. 

At low interaction strength, the system is in a superfluid phase for any value of the tilt angle $\theta$. The superfluid phase is destabilized towards a solid phase as the interaction is increased above a critical value $V_c(\theta)/J$. Filled squares mark the onset of the CB solid, which forms at lower $\theta$, while filled triangles mark the onset of the stripe phase which appears at larger $\theta$. 

Using energy considerations, it is easy to see that the anisotropic nature of the dipolar interaction leads to the stabilization of the CB solid at $\theta=0$ and the SS phase at $\theta=\pi/2$. Moreover we observe that at larger values of the tilt angle the superfluid phase is less stable against increasing the dipolar interaction strength. Indeed, at a tilt angle $\theta=\pi/2$, the dipolar interaction strength needed to destroy superfluidity in favor of a solid is about a factor of two smaller than what needed at a tilt angle $\theta=0$. At intermediate tilt angles, $\theta\sim\pi/6$, the competition between the two solid orders renders both unstable. This adds to the robustness of the superfluid phase, where a dipolar interaction $V/J\sim 6$ is required in order to destroy the off-diagonal long range order.

We find the solid-SF transition points by studying finite size effects on $\rho_{s}$ and $S(\mathbf{k})$. We determine the error bars for the solid to superfluid  transition by monitoring the order parameters characterizing the two phases, namely $\rho_s$ and $S(\mathbf k)$. We assure that the order parameter characterizing the solid phase vanishes with increasing system size as we cross the phase boundary to the superfluid phase and vice versa.

\begin{figure}[h]
\includegraphics[trim={1cm 4cm 14cm 2cm}, clip, width=0.5\textwidth]{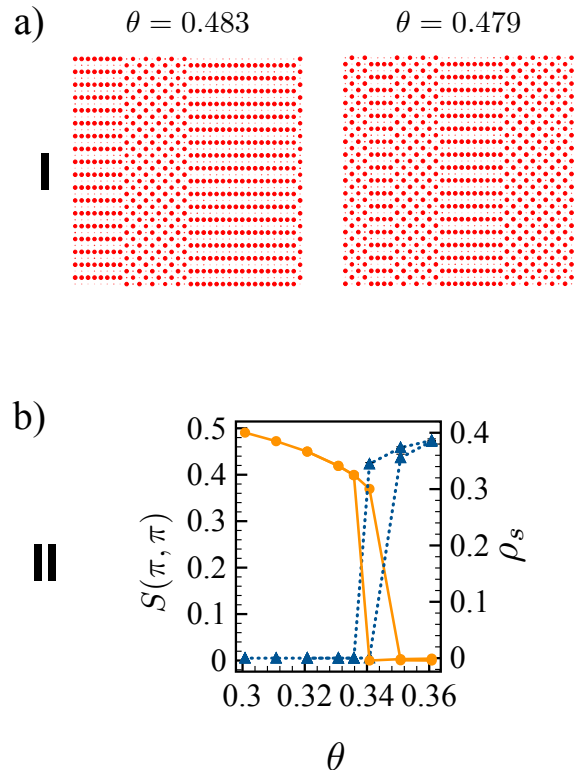}
\caption{(Color online) 
The behavior of the system along cuts I and II. (a) The density distribution of the system along cut I through the shaded region in figure~\ref{fig:fig2} at two values of $\theta$ shows evidence of micro emulsion phase. Here, each circle corresponds to a different site, and its radius is proportional to the local density. (b) The CB structure factor $S(\pi,\pi)$ and superfluid density $\rho_s$ as a function of $\theta$ along cut II. We find evidence of hysteresis which indicates that this transition behaves as a weak first-order phase transition for finite systems. }
 \label{fig:fig2b}
\end{figure}

Given the nature of the interaction and the dimensionality of the system, a first order phase transition between two ordered states is not be allowed~\cite{Ng:1995ca,Spivak:2003ka,Spivak:2004jk,Pollet:2010cc} and a second order phase transition is highly unlikely. %It should be noted that we see no evidence of a supersolid phase appearing between the solid and the superfluid phases. 
First we investigate the nature of the phase transition between the CB and SS phases $V/J=8$. To this end we perform scans along the cut marked I on the main panel of figure~\ref{fig:fig2}. We find that the CB-SS phase transition at $T=0$ is neither first-order nor second order and follows the Spivak-Kivelson micro-emulsion scenario as predicted in~\cite{Ng:1995ca,Spivak:2003ka,Spivak:2004jk,Pollet:2010cc} which is indicated by the shaded region in Fig.~\ref{fig:fig2}. We provide examples of the density map of the system in this region in figure~\ref{fig:fig2b}(a). Here, each circle corresponds to a different site, and its radius
is proportional to the local density. In this region the system features a micro-emulsion phase, where stripes of CB phase are embedded in the SS phase or vice versa. 

We have also investigated the nature of the CB-SF phase transition by performing a similar scan along $V/J=4$ line labeled as II, for a system with $L=100$. In figure~\ref{fig:fig2b}(b) we plot the structure factor $S(\pi,\pi)$ as a function of $\theta$ using filled circles. The superfluid density $\rho_s$ is shown using filled triangles. We observe hysteresis in both $S(\pi,\pi)$ and $\rho_s$ with a width of the hysteresis loop of only $\Delta\theta\sim 0.015$. While this observation suggests that the system undergoes a weak first-order phase transition, this is not consistent with theoretical predictions stated above. Moreover our inability to observe a micro emulsion phase intervening between the superfluid and the CB phase for system sizes up to $L=100$ suggests that we may be far away from the thermodynamic limit. Hence if the micro-emulsion phase exists, the size of emulsion would be larger than this lattice size, and correspondingly large systems should be used in order to observe it experimentally.

%\emph{Finite temperature results}

\begin{figure}[h]
\includegraphics[trim={1cm 2cm 2cm 1cm}, clip, width=0.5\textwidth]{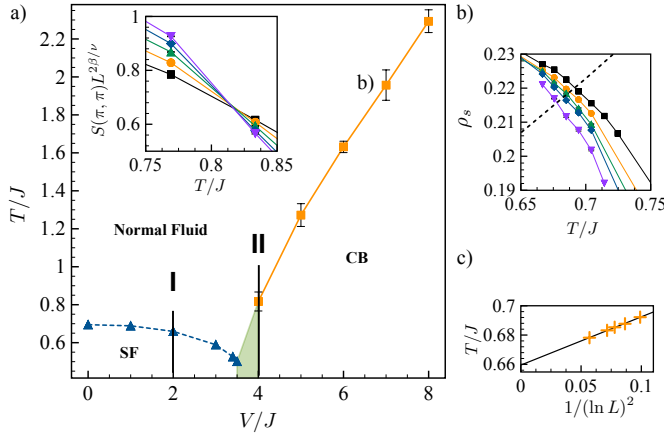}
\caption{(Color online)(a) Finite temperature phase diagram at tilt angle $\theta=0.271$. Upon increasing the temperature, thermal fluctuations destroy the CB and SF order in favor of a normal fluid. The CB phase melts via a two-dimensional Ising transition. The inset shows the scaled structure factor with $2\beta/\nu=0.25$ for $L=18$, 24, 36, 42, 48 along cut II, at $V/J=4$. The crossing determines the critical temperature $T_c/J=0.812\pm 0.002$. The SF-normal fluid transition is a Kosterlitz-Thouless phase transition. In (b) we show $\rho_s$ as a function of $T/J$ for $L=18$, 24, 36, 42, 48, and 66 along cut I at $V/J=2$. The dashed line is given by $T/\pi$. In (c) we use the intersection points between the $T/\pi$ line and the $\rho_s$ vs. $T/J$ curves at each $L$ to extract $T_c/J\approx 0.66$. }
\label{fig:fig3}
\end{figure}

Next we present an investigation of the robustness of the quantum phases described above against thermal fluctuations. As expected, we find the solid phases to be the most robust against thermal fluctuations. We have performed scans over $V/J$ at two tilt angles $\theta=0.271$ and $\theta=0.84$ corresponding to SF-CB and SF-SS transitions at zero temperature. Our results for $\theta=0.271$ are summarized in figure~\ref{fig:fig3}. The phase boundaries are extracted from finite size scaling analysis of $\rho_s$ and $S(\pi,\pi)$. We were not able to resolve the phase boundaries within the shaded region with the system sizes considered in this paper. Within this region we expect the system to undergo either a weak first order phase transition or feature a micro-emulsion phase at zero temperature. Upon increasing the temperature, thermal fluctuations destroy the SF phase in favor of a normal fluid via a Kosterlitz-Thouless (KT) transition~\cite{Kosterlitz:1973fc}. In figure~\ref{fig:fig3}(b) we show $\rho_s$ as a function of $T/J$ for $L=18$, 24, 36, 42, 48, and 66 at $V/J=2$, indicated as cut I in figure~\ref{fig:fig3}(a). In the thermodynamic limit, a universal jump is observed at a the critical temperature given by $\rho_s=2m k_B T_c/\pi \hbar^{2}$. In a finite size system this jump is smeared out as seen in figure~\ref{fig:fig3}(b). We extract the critical temperature using the finite size scaling procedure described in~\cite{Ceperley:1989hb}. The dashed line in figure~\ref{fig:fig3}(b) corresponds to $\rho_s=T/\pi$ and its intersection points with each $\rho_s$ vs. $T/J$ curve are used to find $T_c$ as shown in figure~\ref{fig:fig3}(c). We find $T_c/J\approx 0.66$ at $V/J=2$. 
The CB solid melts in favor of a normal fluid via a two-dimensional Ising transition.  We use standard finite size scaling as shown in the inset of figure~\ref{fig:fig3} where we plot the scaled structure factor $S(\pi,\pi)L^{2\beta/\nu}$, with $2\beta/\nu=0.25$  as a function of $T/J$ for $L=18$, 24, 36, 42, 48 along cut II, at $V/J=4$. The crossing indicates a critical temperature $T_c/J=0.812\pm 0.002$.

\begin{figure}[h]
\includegraphics[trim={1cm 8cm 4cm 1cm}, clip, width=0.5\textwidth]{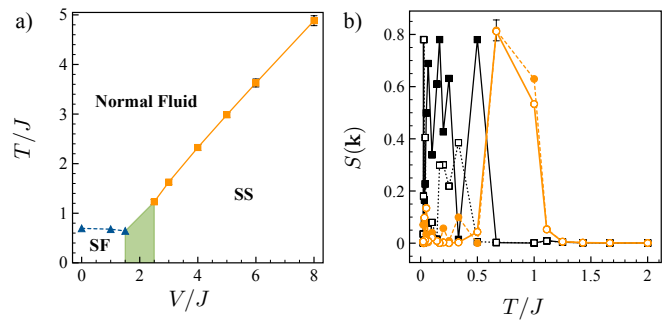}
\caption{(Color online)(a) Finite temperature phase diagram at tilt angle $\theta=0.84$. 
%Upon increasing the temperature thermal fluctuations destroy the SS and SF order in favor of a normal fluid.
Critical temperatures are found using the same methods as described for the data in figure~\ref{fig:fig3}. (b) Finite temperature properties of the CB-SS micro-emulsion phase at $V/J=8$, $\theta\approx 0.48$, and $L=100$. Squares and circles correspond to $S(\pi,\pi)$ and  $S(\pi,0)$, respectively. Filled and empty symbols refer to different initial conditions. At $T/J \gtrsim 1.2$ the system is in the normal fluid phase, while at $T/J<1.2$, the micro-emulsion phase is dominated by either solid order, depending on the initial conditions. }\label{fig:fig4}
\end{figure}

We have performed a similar analysis at fixed tilt angle $\theta=0.84$ where the zero-temperature phase diagram features the SF and SS phases. The results are shown in figure~\ref{fig:fig4}(a). Critical temperatures are found using the same methods as described for the SF and CB phases. It is worth noting that while both the SS and CB phases are more robust against thermal fluctuations compared to the SF phase, at any given $V/J$ the SS phase has critical temperature roughly twice that of the CB phase.  

Finally, we look at the finite temperature properties of the CB-SS micro-emulsion phase at $V/J=8$. In figure~\ref{fig:fig4}(b) we show $S(\mathbf{k})$ for the CB and SS phases as a function of $T/J$ at $\theta\approx 0.48$ and $L=100$.  We use squares and circles to show $S(\pi,\pi)$ and  $S(\pi,0)$, respectively. The filled and open symbols correspond to two different initial conditions chosen for the simulations at each $T/J$. Since the equilibrium density distribution of the micro-emulsion phase is affected by the choice of initial conditions we observe large fluctuations in $S(\mathbf{k})$. These results are in agreement with what reported in~\cite{Pollet:2010cc}. We find that at $T/J \gtrsim 1.2$ the system is in the normal fluid phase, with both SS and CB phases disappearing. If $T/J<1.2$ the micro-emulsion phase can be dominated by either solid order, depending on the initial conditions.

\label{sec:sec3}\section{Harmonic confinement}
In a typical experimental system particles are subject to the optical lattice potential as well as an external harmonic confinement. The effect of the harmonic confinement can be taken into account in the QMC simulations through a site-dependent chemical potential as shown in Hamiltonian~\eqref{eq:H}. 

\begin{figure}[h]
\includegraphics[trim={1cm 1cm 5cm 0cm}, clip, width=0.5\textwidth]{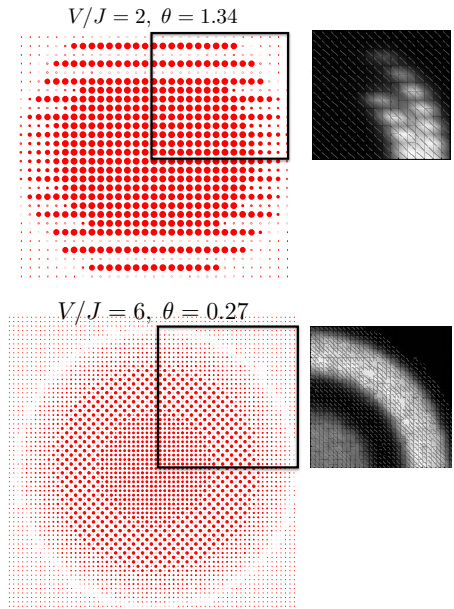}
\caption{(Color online) Left panels: Density distribution of dipoles in the presence of a harmonic potential (see text). Each circle corresponds to a different site, and its radius is proportional to the local density.  Right panels: Statistics of the off-diagonal correlator $\langle a^\dagger(\vec r_i) a(\vec r_j)\rangle$ (see text for detail). A brighter color corresponds to regions of the trap where off-diagonal long-range order is present. Top left and right panels: $V/J=2$, $\theta=1.34$, $N\sim 511$. A Mott insulator phase with filling factor one is observed at the center of the trap, while a stripe solid is stabilized in most of the outer shell. Bottom left and right panels: $V/J=6$, $\theta=0.27$, $N\sim 1217$. A superfluid phase is observed at the center of the trap, followed by a shell of CB solid which gives way to a superfluid outer shell once the harmonic potential is strong enough destroy the solid order.}\label{fig:fig5}
\end{figure}
Experimentally, using a sufficiently deep optical lattice suppresses double occupancy. In this limit, the hard-core condition is valid for all tilt angles. The variation of the chemical potential provides a scan over density hence resulting in coexistence of phases realized at half-filling (as described above) with phases stabilized at other fillings. We use a lattice depth $V_0=50E_R$, where $E_R$ is the recoil energy of the molecule and a harmonic confinement of $\omega_{x,y}=2\pi \, 5$~Hz. Our parameters reflect typical experimental choices. The top and bottom left panels in figure~\ref{fig:fig5} present the equilibrium density distribution of the dipoles, where each circle corresponds to a different site, and its radius
is proportional to the local density. 
In the presence of the harmonic trapping potential we cannot use the usual winding number relations to measure the superfluid density in the different regions of the trap. However we can use the off-diagonal correlator $\langle a^\dagger(\vec r_i) a(\vec r_j)\rangle$, to map the superfluid regions within the trap. The panels on the right-hand side of figure~\ref{fig:fig5} show the regions in the trap where the correlator is long-ranged. Each panel is normalized such that the brighter points are those exhibiting  long-ranged  correlations and thus robust superfluidity. 

The top left and right panels correspond to $V/J=2$, tilt angle $\theta=1.34$ and $N\sim 511$. The particles at the center of the trap form a Mott insulator (MI) with unit filling while a stripe solid is stabilized in most of the outer shell. The dark regions in the top right panel indicate an absence of off-diagonal long-range order, corresponding to the SS and MI phases. Given the steepness of the harmonic potential, we do not observe a superfluid shell separating the two insulating phases, however off-diagonal long-range order is present adjacent to the MI shell. This is due to the mismatch between the symmetry of the SS and the trap, which destabilizes the SS and allows for the build up of phase coherence in this region. 

The bottom left and right panels of figure~\ref{fig:fig5} corresponds to $V/J=6$ and tilt angle $\theta=0.27$ for which the CB phase is stabilized at half filling. The total number of particles in the trap is $N\sim 1217$. We observe the presence of superfluid phase at the center of the trap, followed by a shell of CB solid which gives way to a superfluid outer shell once the harmonic potential is strong enough destroy the solid order. The superfluid regions are characterized by local density different than $1$ and $0.5$ as the left panel shows, while off-diagonal long-range order is present as seen in the right panel. On the other hand, the left panel clearly shows a CB structure at density $0.5$ in the inner shell with corresponding absence of off-diagonal long-range order (dark color in right panel).

\label{sec:sec4}\section{Experimental Realization}

The system described in section~\ref{sec:sec1} can be experimentally realized using bosonic polar molecules trapped in a two-dimensional lattice, provided that the lattice is deep enough to suppress double occupancy hence ensuring the hard-core condition. Once ground-state molecules are formed in the lattice their dipole moments can be aligned through the application of an external static electric field, realizing an interaction term of the form described in Eq.~\eqref{eq:H}. While equilibrium states such as the ones described above have not been prepared yet, rapid experimental advances will allow for the realization of these states in the future. Our calculations suggest that, unless dipoles are tilted at intermediate angles $\theta\sim\pi/6$, for which superfluidity survives even for suppressed hopping strength $J/V\sim 0.17 $, solid phases would be the best candidate for experimental observation since they are more robust against thermal fluctuations. For example, for a system of $^{39}$K$^{87}$Rb polar molecules  with permanent dipole moment $d=0.57D$, trapped in an optical lattice with lattice constant $a=532$ nm and depth $V_0=40E_{r}$, with tunneling rate $J/h=104$ Hz~\cite{Yan:2013fn}, %and $V/h=338$ Hz% 
one can achieve $J/V\sim 0.3$. Hence, observation of the stripe solid would be possible at temperatures $\sim10$ nK. With the same experimental setup the emulsion phase can be observed at temperatures $\sim6$ nK. If we consider $^{23}$Na$^{40}$K with permanent  dipole moment $d=2.7D$~\cite{Park:2015vz} and same trapping parameters as above, one can achieve $J/V \sim 0.0136$ which allow for observation of the CB solid phase and the SS phase at temperatures $\sim109$ nK and $\sim250$ nK respectively. It should be noted that similar behavior is expected in experimental systems featuring atoms and molecules with magnetic dipole moments, such as Er$_2$, Cr, and Dy \cite{Griesmaier:2005fd, Frisch:2015wu,Lu:2012bd}. 

\label{sec:sec5}\section{Conclusions}
In summary, we presented the zero- and finite-temperature phase diagrams of a system of hard-core, dipolar lattice bosons at half-filling as a function of the alignment of the dipole moments, characterized by the tilt-angle $\theta$, and the strength of the dipolar interaction. At zero-temperature, the system features three phases: superfluid, checkerboard solid and stripe solid. The superfluid phase is present at all tilt angles, provided that the interaction strength is below $V_c(\theta)/J$, and upon increasing the interaction strength the system enters one of the two solid phases depending on the value of $\theta$. We observed signatures of a micro-emulsion phase transition, observing an emulsion phase of the checkerboard and stripe solids in the region separating the two solid phases. We expect solid phases to be stabilized by lattice dipolar fermions as well as shown in~\cite{Bhongale:2012fb}. Finally, we studied the robustness of these phases against thermal fluctuations and showed that the solid phases can be observed experimentally at temperatures up to $\sim 100$~nK. 
\section{Aknowledgments}
This work was supported by the NSF (PIF-1415561, PIF-1211914 and PFC- 1125844), AFOSR, AFOSR-MURI, NIST and ARO individual investigator awards. The computation for this project was performed at the OU Supercomputing Center for Education and Research (OSCER) at the University of Oklahoma (OU).

\bibliography{tilt}

%\begin{thebibliography}{50}
%\bibitem{Santos} X. Deng, R. Citro, E. Orignac, A. Minguzzi, and L. Santos\njp{15}{045023}{2013}.
%\end{thebibliography}

\end{document}